\documentclass[12pt,epsfig]{article}
\usepackage[dvips]{epsfig}
\psfigdriver{dvips}

\begin{document}

%%%%%%%%%%%%%%%%%%%%%%%%%%%%%%%%%%%
%

\def\papertitlepage{\baselineskip 3.5ex \thispagestyle{empty}}
\def\preprinumber#1#2{\hfill \begin{minipage}{4.2cm}  #1
                 \par\noindent #2 \end{minipage}}
\renewcommand{\thefootnote}{\fnsymbol{footnote}}
\newcommand{\beq}{\begin{equation}}
\newcommand{\eeq}{\end{equation}}
\newcommand{\beqa}{\begin{eqnarray}}
\newcommand{\eeqa}{\end{eqnarray}}

%
%%%%%%%%%%%%%%%%%%%%%%%%%%%%%%%%%%%%%%%%%%%%%%%
% 
\papertitlepage
\setcounter{page}{0}
\preprinumber{KEK Preprint 2003-52}{hep-th/0309049}
\baselineskip 0.8cm
\vspace*{2.0cm}
\begin{center}
{\large\bf Remarks on Phase Transitions in Matrix Models \\ 
and ${\cal N}=1$ Supersymmetric Gauge Theory}
\end{center}
\vskip 4ex
\baselineskip 1.0cm
\begin{center}
           {Hiroyuki~ Fuji\footnote[1]{\tt hfuji@post.kek.jp} 
 and Shun'ya~ Mizoguchi\footnote[2]{\tt mizoguch@post.kek.jp} } 
\\
    \vskip -1ex
       {\it High Energy Accelerator Research Organization (KEK)} \\
    \vskip -2ex
       {\it Tsukuba, Ibaraki 305-0801, Japan} \\
% \vskip 1ex
%
\end{center}
%%%%%%%%%%%%%%%%%%%%%%%
\vskip 10ex
%%%%%%%%%%%%%%%%%%%%%%%
%
\baselineskip=3.5ex
\begin{center} {\large\bf Abstract} \end{center}
A hermitian one-matrix model with an even quartic potential exhibits 
a third-order phase transition when the cuts of the matrix model curve 
coalesce. We use the known solutions of this matrix model to compute 
effective superpotentials of  an ${\cal N}=1$, $SU(N)$ supersymmetric 
Yang-Mills theory coupled to an adjoint superfield, following the techniques 
developed by Dijkgraaf and Vafa. These solutions automatically satisfy the 
quantum tracelessness condition and describe a breaking to 
$SU(N/2)\times SU(N/2) \times U(1)$. We show that the value of the 
effective superpotential is smooth at the transition point, and that 
the two-cut (broken) phase is more favored than the one-cut (unbroken) 
phase below the critical scale.  The $U(1)$ coupling constant diverges 
due to the massless monopole, thereby demonstrating Ferrari's general formula. 
We also briefly discuss the implication of the Painlev\'e II equation arising 
in the double scaling limit.
 
\vskip 2ex
\vspace*{\fill}
\noindent
September 2003
\newpage
\renewcommand{\thefootnote}{\arabic{footnote}}
\setcounter{footnote}{0}
\setcounter{section}{0}
\baselineskip = 0.6cm
\pagestyle{plain}

\section{Introduction}
Over the recent couple of years, much progress has been made in computing 
effective superpotentials in ${\cal N}=1$ supersymmetric gauge theories. The 
idea for this technique was motivated by geometric considerations of 
dualities in string theory \cite{CIV,CV}, and then it was recognized that 
the computation was closely related to that in the old matrix 
models \cite{DV1,DV2,DV3}. Later the conjecture was proved under some 
certain conditions \cite{DGLVZ,CDSW}, 
and the validity of this approach has been extensively tested in a variety 
of situations.

This approach provides direct connections between the computations 
in the old matrix models to those in supersymmetric gauge theories,  
enabling us to `recycle' \cite{SeibergKyoto} the old matrix model results 
to extract interesting information on gauge theory dynamics from them, 
without doing many new computations. 
In this paper, we use some known solutions of a matrix model that exhibits 
a phase transition, and examine what it implies to the corresponding gauge 
theory. 

Our model is a hermitian one-matrix model with a familiar symmetric quartic 
potential
\beqa
W_{\mbox{\scriptsize tree}}(M)&=&\frac m2 M^2 + \frac{g_4}4 M^4.
\label{Wtree}
\eeqa
The one-cut solution of this model is well-known \cite{BIPZ}, and a two-cut 
solution was also obtained in \cite{Shimamune} (See also \cite{CMM}.). 
Many things are known on this model; for example, it exhibits a third-order 
one-cut/two-cut phase transition \cite{Shimamune}, and in the double scaling 
limit the critical behavior is described by the Painlev\'e II 
equation \cite{DSS}. The same equation was found to appear \cite{PS} near 
the Gross-Witten transition point \cite{GW} of a unitary matrix model.  

Following the techniques recently developed by Dijkgraaf and Vafa, we will 
use these solutions to compute low-energy effective superpotentials for 
${\cal N}=1$, $SU(N)$ supersymmetric gauge theories, for both the phase  
with the maximally unbroken gauge group and that broken to 
$SU(N/2)\times SU(N/2) \times U(1)$ for an even $N$. Since the solutions
are all $Z_2$ symmetric, the matrix models automatically satisfy the quantum 
tracelessness condition \cite{CIV}, and hence describes this particular 
pattern of gauge symmetry breaking.

We will show that the values of  the effective superpotentials are smoothly 
connected at the transition point; the two-cut value of the superpotential is 
lower than that of the one-cut case below the critical scale, being consistent 
with what one would naively expect from the renormalization group argument.
We will also confirm Ferrari's general formula \cite{Ferrari2} for the 
critical behavior of the U(1) coupling constant.  At the transition point, 
the $U(1)$ coupling constant diverges, signaling the effect of other 
degrees of freedom (the monopole), and  the $U(1)$ kinetic term consistently 
disappears there from the effective action.  We will also briefly discuss 
what can be learned about the gauge theory from the Painlev\'e II equation 
of the double-scaled matrix model.

We would like to emphasize what is new in this paper compared to the earlier 
works, in particular Ref.\cite{Ferrari2}. Our main focus is on the use of 
classic $Z_2$ symmetric solutions of old matrix models and their $SU(N)$ 
(rather than $U(N)$) gauge theoretical interpretation, by means of  
Dijkgraaf and Vafa's procedure.  The fact that the smoothness of 
the `on-shell' value of the effective superpotential has already been 
mentioned in \cite{Ferrari2,Shih}, but the comparison of the values below 
the critical scale has not been done. Ferrari \cite{Ferrari2} 
has also presented a nice derivation of the general U(1) 
coupling constant formula, and our analysis using Shimamune's 
matrix model curve will be 
a check for this. We will also give its explicit off-shell form; 
in fact, it is the extra double-zero factor that enables us to analytically 
solve the model in terms of elementary functions. The discussion on the 
implication of the Painlev\'e II will also be novel.
  
The matrix model solution we focus on in this paper can be thought of as a 
singular limit of a three-cut solution, although this curve itself cannot be 
realized as an ${\cal N}=2$ Seiberg-Witten curve \cite{SW, KLYT,AF} with a 
finite $\Lambda$. Some ${\cal N}=1$ gauge dynamics near the 
Argyres-Douglas like singularities \cite{AD} has been studied in 
\cite{CSW,ES,B,Shih}.
 
\section{The one-cut versus two-cut solutions in a hermitian one-matrix model}
We consider a hermitian one-matrix model with an even quartic tree-level 
potential (\ref{Wtree})
with an $\hat N\times \hat N$ hermitian matrix $M$. The free energy is 
given by
\beqa
F&=&-\log\int dM \exp\left(-\frac{\hat N}\mu \mbox{Tr} 
W_{\mbox{\scriptsize tree}}(M)\right)
\nonumber\\
&=&\sum_{g=0}^\infty \hat N^{-2g +2}F_g,
\eeqa
where we have included the matrix model 't Hooft coupling $\mu$. 
Due to the redundancy of parameters, we will set $g_4=1$ 
in the following; clearly, the $g_4$-dependence can be recovered by 
replacing $\mu \rightarrow \frac \mu{g_4}$ and $m \rightarrow 
\frac m{g_4}$.  $\mu>0$ is also assumed in this paper.

\subsection{The one-cut solution}
The BIPZ large-$\hat N$ one-cut free energy is given \cite{BIPZ} in our 
notation
\beqa
F_0&=&-\frac1{24}(A-1)(A-9)
-\frac12\log\frac{\mu A}{m\Lambda_0^2}+\frac34 
\eeqa
with
\beq
A= \frac{m b_1^2}{4\mu}, ~~~
b_1^2=\frac23 (\sqrt{12\mu-m^2}-m).
\label{b(1-cut)}
\eeq
$\Lambda_0$ is an arbitrary integration constant and can be identified 
as the cutoff parameter in the corresponding gauge theory 
(See next section.). In the original BIPZ formula, a constant 
\beq
F_0 (g_4\!=\!0)
=-\frac12\log\frac \mu{m\Lambda_0^2}+\frac34 \label{F(g=0)}
\eeq
is subtracted from the free energy ($g_4\!=\!0$ implies $A\!=\!1$.).

The resolvent $\omega(z)=\ \frac1{\hat N}\mbox{Tr}\frac1{z-M}$
and the spectral density $\rho(\lambda)= -\frac1{2 \pi i}
(\omega(\lambda+i 0)-\omega(\lambda-i0))$ are
\beqa
\omega(z)&=&\frac1{2\mu}\left(
mz+z^3-(m+\frac{b_1^2}2+z^2)(z^2-b_1^2)^{\frac12}
\right),
\label{omega(1-cut)}
\\
\rho(\lambda)&=&\frac1{2\pi \mu}(m+\frac{b_1^2}2+\lambda^2)
\sqrt{b_1^2-\lambda^2}.
\eeqa
$\lambda$, the eigenvalue of $M$, is distributed only on the interval 
between $-b_1$ and $b_1$. 
$\rho(\lambda)=0$ otherwise.

If $m>0$,  $\rho(\lambda)$ 
is always positive on 
the interval $[-b_1,b_1]$, while if $m$ is negative and 
$m<-2\sqrt{\mu}$, $\rho(\lambda)$ takes a negative value 
in a region near $\lambda=0$. This negative eigenvalue density is 
unacceptable as a matrix model, indicating a split of the cut. 

\subsection{The two-cut solution}
The symmetric two-cut solution of this matrix model was obtained by Shimamune 
\cite{Shimamune}
(See also \cite{CMM}.). The free energy is
\beqa
F_0=-\frac14 \log\frac \mu{\Lambda^4}-\frac{m^2}{4\mu}+\frac38.
\label{F0(2-cut)}
\eeqa
Again, to compare with the literature, 
the constant (\ref{F(g=0)}) needs to be subtracted. 
The resolvent and spectral density are
\beqa
\omega(z)&=&
\frac1{2\mu}\left(mz + z^3 - z(z^2-a^2)^{\frac12}(z^2-b^2)^{\frac12}\right),
\label{omega(2-cut)}
\\
\rho(\lambda)&=&
\frac1{2\pi \mu}|\lambda|\sqrt{(\lambda^2-a^2)(b^2-\lambda^2)},
\eeqa
where
\beq
a^2=-m-2\sqrt{\mu},~~~b^2=-m+2\sqrt{\mu}. 
\label{ab}
\eeq
The eigenvalues are symmetrically distributed on the two 
intervals ${[}\!-\!b,-a{]}$ and ${[}a,b{]}$.

In deriving (\ref{F0(2-cut)}), some care 
must be taken because there is no eigenvalue at $\lambda=0$ for the two-cut 
solution, and therefore the integrated saddle point equation 
\beq
\int d\lambda' \frac{\rho(\lambda')}{\lambda-\lambda'}
=\frac1{2\mu} W'(\lambda)
\eeq
does not hold at $\lambda=0$. Instead, taking $\lambda$ to be an end 
of a cut, one obtains
\beqa
F_0&=&
\int_a^b d\lambda \rho(\lambda)
\left(
\frac1\mu W(\lambda)-\log|\lambda+a|-\log|\lambda-a|
\right)+\frac1{2\mu}W(a),
\eeqa 
which replaces the free-energy formula of \cite{BIPZ}.

For $\mu$ fixed, the two cuts get closer as $| m |$ decreases from a large 
negative value,  until $m$ reaches to $- 2\sqrt\mu$ when the two end points 
coalesce. One sees that  the third derivative of the free energy 
with respect to the couplings is discontinuous at this point, exhibiting a
third-order phase transition \cite{Shimamune,CMM}. 

\section{Phase Transition in Gauge Theory}
We will now use the results in the previous section and study what they 
mean in the corresponding 4d ${\cal N}=1$
supersymmetric gauge theory coupled to an adjoint chiral superfield $\Phi$ 
with the same (super)potential 
\beq
W_{\mbox{\scriptsize tree}}(\Phi)= \frac m2\Phi^2 + \frac 14 \Phi^4. 
\eeq 
As we mentioned in the introduction, the solutions in the previous section 
describe gauge theories with gauge groups $SU(N)$ and 
$SU(N/2) \times SU(N/2) \times U(1)$ in the classical vacua 
(Therefore when we talk about the transition, $N$ must be even.), 
respectively.

The procedure to compute the effective superpotential is summarized as 
follows \cite{CIV, DV1,DV2,DV3,CDSW}. Each branch cut of a resolvent 
$\omega(z)$ corresponds to a non-abelian factor of the unbroken gauge 
group. Let $n$ be the number of cuts, and (initially) the gauge group $U(N)$ 
be partially broken to $\prod_{i=1}^n SU(N_i) \times U(1)^n$ 
($\sum_{i=1}^n N_i = N$).  
One then computes the low-energy effective superpotential 
$W_{\mbox{\scriptsize eff}}$ as 
\beqa
W_{\mbox{\scriptsize eff}}
&=&2\pi \sum_{i=1}^n ( i \tau S_i +  N_i \Pi_i)
+\frac12\sum_{i,j=1}^n \frac{\partial \Pi_i}{\partial S_j}
w_{\alpha i}w_j^\alpha,
\label{Weffdef}
\eeqa
where $S_i$ and $w_{\alpha i}$ are the glueball and the $U(1)$ 
superfields 
\beqa
S_i&=&-\frac1{32\pi^2}\mbox{Tr}W_{i\alpha}W_i^\alpha,
~~~w_{i\alpha}~=~\frac1{4\pi}\mbox{Tr}W_{i\alpha}
\eeqa
for each $U(N_i)$ factor. $S_i$ and $\Pi_i$ can be 
computed as a period of a complex curve
\beqa
y&=&2\mu\omega(z)-W'_{\mbox{\scriptsize tree}}(z).
\eeqa
The contour for $S_i$ surrounds the $i$-th cut, while for $\Pi_i$ starts from 
the cutoff $\Lambda_0$ on the second sheet, goes through the cut and back 
to $\Lambda_0$ on the first sheet. 
$\tau$ is the
$\Lambda_0$-dependent 
bare coupling constant. The formula 
(\ref{Weffdef}) is for $U(N)$ gauge theories, but separating 
the $SU(N_i)$ piece from $S_i$ as
\beqa
S_i&=&\hat{S}_i-\frac1{2 N_i}w_{\alpha i}w_i^\alpha, 
\eeqa  
the  overall $U(1)$ contributions cancel out, leaving only
the bare coupling term \cite{CDSW}. Discarding this, one obtains an 
$SU(N)$ effective superpotential.

\subsection{The one-cut case}
First, we consider the one-cut solution.
This corresponds to the case when the full gauge group ($SU(N)$) is 
unbroken and maximally confined at low energies, and was already 
studied in e.g.
\cite{Ferrari1,FO,CDSW}. The matrix model curve 
\beqa
y(z)&=&
-(z^2+m+{\textstyle \frac{b_1^2}2})(z^2-b_1^2)^{\frac12}
\label{1-cutcurve}
\eeqa
has a branch cut at ${[}\!-\!b_1,~b_1 {]}$ and two zeroes at 
$z=\pm\sqrt{-m-\frac{b_1^2}2}$.
If $m$ is positive and very large compared to a fixed $\mu$, these zeros 
are pure imaginary and far apart from each other, and from the real axis. 
They move toward $z=0$ as $m$ decreases, but they are still off 
the real axis when $m=0$ (where the tree-level potential develops 
a quadruple zero), until they meet at $z=0 $ when $m$ reaches 
$-2\sqrt{\mu}$.  This is the critical point discussed in section 2. 
$m<-2\sqrt{\mu}$ is the region where the two-cut solution 
is well-defined, but formally a one-cut solution still exists, although the 
spectral density $\rho(\lambda)$ becomes not positive definite. In this 
region the two zeroes are apart, located on the real axis, and 
$\rho(\lambda)$ is negative between them (Figure 1). 

\begin{center}
\begin{figure}[h!]
\begin{tabular}{c}

\begin{minipage}{140mm}
\begin{center}
\makebox{$W_{\mbox{\scriptsize tree}}(\Phi)$}\\
\vspace{2mm}
%\rotatebox{90}{\hspace{3cm}$W_{\mbox{\scriptsize tree}}(\Phi)$}
\resizebox{!}{5cm}{
\includegraphics[width=50mm]{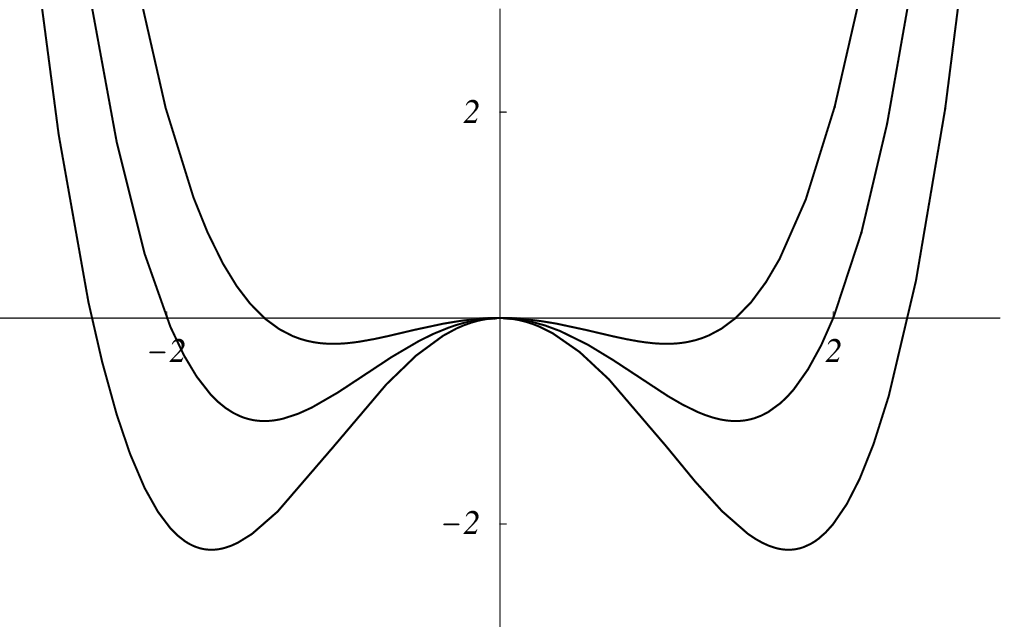}
}
\raisebox{24mm}{$\Phi$}
\\\makebox{\phantom{}\hspace{-4mm}(a)}
\end{center}
\end{minipage}\vspace{5mm }\\

\begin{tabular}{ll}
\begin{minipage}{40mm}
\begin{center}

\resizebox{!}{35mm}{\phantom{}\hspace{-10mm}
\includegraphics[width=35mm]{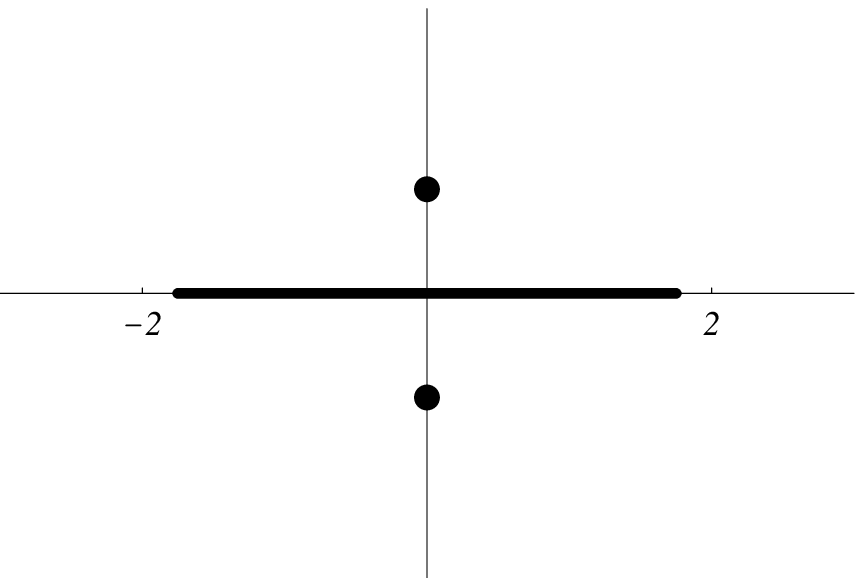}
}
\mbox{\phantom{}\hspace{-13mm}(b)}
\end{center}  
\end{minipage}
&
\begin{minipage}{40mm}
\begin{center}

\resizebox{!}{35mm}{\phantom{}\hspace{5mm}
\includegraphics[width=35mm]{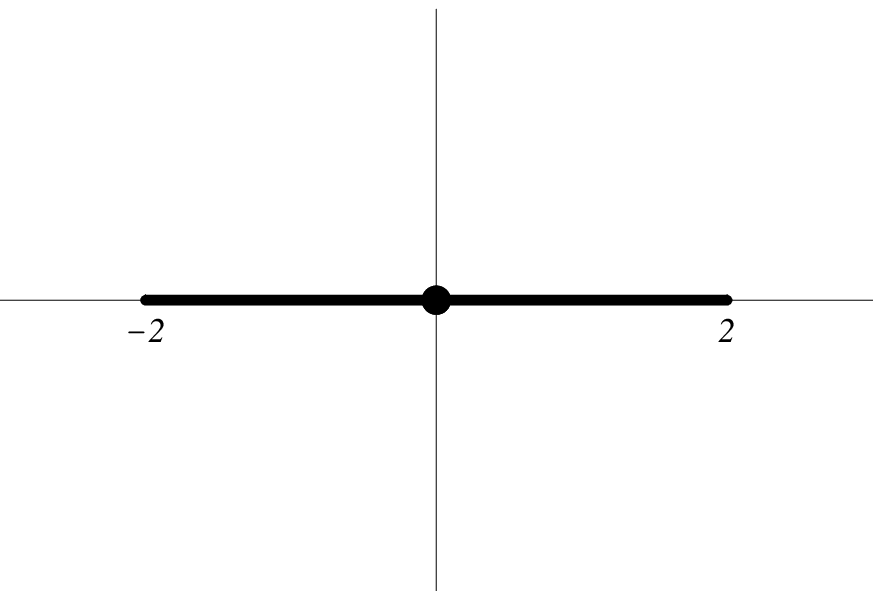}
}
\mbox{\phantom{}\hspace{32mm}(c)}
\end{center}
\end{minipage}
\\
\\
\begin{minipage}{40mm}
\begin{center}

\resizebox{!}{35mm}{\phantom{}\hspace{-10mm}
\includegraphics[width=35mm]{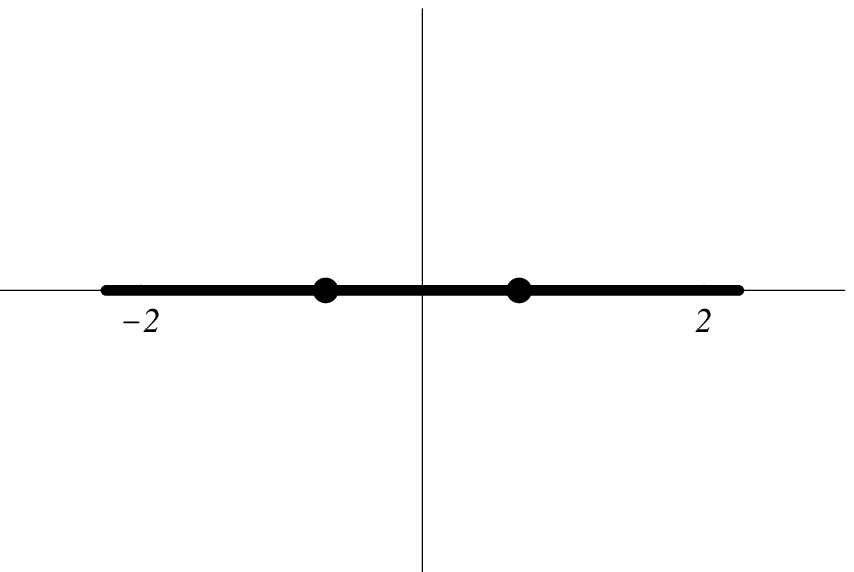}
}
\mbox{\phantom{}\hspace{-13mm}(d)}
\end{center}
\end{minipage}
&
\begin{minipage}{40mm}
\begin{center}

\resizebox{!}{35mm}{\phantom{}\hspace{5mm}
\includegraphics[width=35mm]{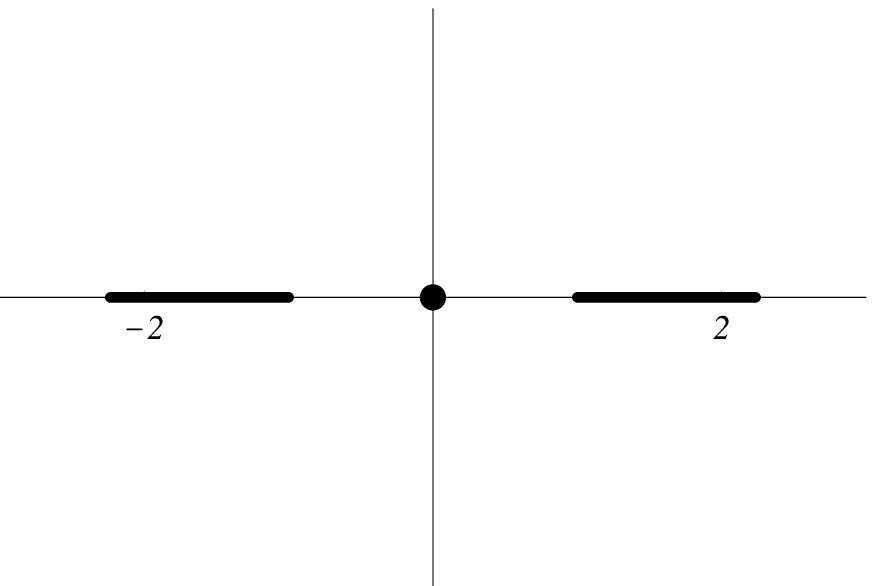}
}
\mbox{\phantom{}\hspace{32mm}(e)}
\end{center}
\end{minipage}

\end{tabular}

\end{tabular}
\vspace{3mm}
\caption{The tree-level superpotentials $W_{\mbox{\scriptsize tree}}$ 
and the cuts and zeroes on the matrix model curves. 
(a) $W_{\mbox{\scriptsize tree}}$ for (from inside) $m=-1$,$-2$ and 
$-3$ with $\mu=1$. 
(b) One-cut, $m=-1$.
(c) One-cut, $m=-2$ (critical).
(d) One-cut, $m=-3$.
(e) Two-cut, $m=-3$.}
\end{figure}
\end{center}

Let $S$ and $\Pi$ denote the unique pair of periods. They are calculated as
\beqa
S&=&\frac1{2\pi i}\int_{-b_1}^{b_1} dz y(z)~=~\mu, 
\label{S=mu}
\\
\Pi&=&\frac1{2\pi}\int_{b_1}^{\Lambda_0} dz y(z)\nonumber\\
&=&-\frac1{2\pi}\left(W_{\mbox{\scriptsize{tree}}}(\Lambda_0)
+2\mu\log\frac{b_1}{2\Lambda_0}-\frac \mu2 -\frac{mb_1^2}8
\right).\label{Pi(1-cut)}
\eeqa
$\Lambda_0$ is the cutoff parameter of dimension $3/4$. 
One can easily verify the special geometry relation 
(with the $\hat N^2$ factor of the genus-0 
free energy taken into account)
$\Pi=\frac{\partial(S^2F_0)}{\partial S}$ 
up to a $\Lambda_0$-dependent constant 
$W_{\mbox{\scriptsize tree}}(\Lambda_0)$. 
Plugging (\ref{S=mu})(\ref{Pi(1-cut)})
into (\ref{Weffdef}) and discarding the overall $U(1)$ term, we obtain \cite{FO}
\beqa
W_{\mbox{\scriptsize eff}}(\hat S)
&=&
N\left(
-2\hat S\log\frac{\hat b}{2\Lambda}+\frac{\hat S}2 +\frac{m\hat b^2}8
\right) 
%-\frac{\pi i\tau}N w_\alpha w^\alpha
~~~\mbox{(one-cut),}
\eeqa
where $\hat b = b_1|_{\mu = \hat S}$.
$\Lambda$ is the physical scale defined through the `renormalization'
of the gauge coupling constant \cite{CIV}
\beqa
\log \Lambda&=&\log \Lambda_0+\frac{\pi i}N \tau.
\label{scales}
\eeqa
$W_{\mbox{\scriptsize eff}}(\hat S)$ is minimized with respect to $\hat S$
at $\hat b = 2\Lambda$ \footnote{with a possible $Z_N$ chiral-symmetry  
phase factor, which we omit in this paper.  Other vacua may be obtained by 
rotating the complex plane on which the matrix model curve is defined.}.
Evaluating $W_{\mbox{\scriptsize eff}}(\hat S)$ at this 
point  reproduces the known $SU(N)$ effective superpotential 
\cite{Ferrari1,CDSW} 
\beqa
W_{\mbox{\scriptsize low}}
&=&
N\left(
\frac{3\Lambda^4}2 
+ m\Lambda^2
\right)
~~~\mbox{(one-cut)}.
\label{Wlow(1-cut)}
\eeqa

\subsection{The two-cut case}
We next consider the two-cut case. 
For general two-cut solutions, (\ref{Weffdef}) computes a superpotential
for the gauge group $U(N)$ broken to 
$SU(N_+) \times SU(N_-)\times U(1)^2$ for arbitrary 
$N_+$ and  $N_- = N-N_+$. On the other hand, our solution 
(\ref{omega(2-cut)}) is $Z_2$ symmetric and the two periods $S_\pm$ are 
not independent (We will denote the period whose contour surrounds the 
cut ${[}a,b{]}$(${[}-b,-a{]}$) by $S_+$($S_-$) and its dual period by 
$\Pi_+$($\Pi_-$).). This constraint can be thought of  
as a consequence of the quantum tracelessness 
condition \cite{CIV}: Suppose that we perturb 
$W_{\mbox{\scriptsize tree}}(\Phi)$ by a small linear potential 
$\delta W_{\mbox{\scriptsize tree}}(\Phi)= \sigma\mbox{Tr}\Phi$. 
Clearly, $\delta\Pi_+ = -\delta\Pi_- = O(\sigma)$, and therefore the symmetric 
solution satisfies the quantum tracelessness condition 
$\frac{\partial W_{\mbox{\scriptsize eff}}}{\partial\sigma}=0$ 
if $N_+=N_-$.  Conversely, if $N_+=N_-$, 
the equation $\frac{\partial W_{\mbox{\scriptsize eff}}}{\partial\sigma}=0$ 
imposes a constraint that  $S_+ =S_-$ and $\Pi_+ = \Pi_-$. Thus our symmetric 
two-cut solution corresponds to a gauge theory with a gauge group 
$SU(N)$ broken to 
$SU(N/2) \times SU(N/2) \times U(1)$ for some even $N$.\footnote{Therefore, 
if $N$ is odd, the gauge theory in the broken phase is not 
described by this symmetric family of matrix model solutions.}

To compute the $U(1)$ coupling in (\ref{Weffdef}) for this phase, 
we still need to have two independent $S_i'$s. For this purpose we slightly 
relax the $Z_2$ symmetry, and seek for a slightly asymmetric solution by a 
perturbation. Let the locations of two cuts be 
${[} b_-,a_-{]}$ and ${[}a_+,b_+{]}$, 
then the resolvent $\omega(z)$ is given by
\beqa
\omega(z)&=&\frac1{4\pi i \mu}
(z-b_-)^{\frac12}(z-a_-)^{\frac12}
(z-a_+)^{\frac12}(z-b_+)^{\frac12}\nonumber\\
&&\cdot
\oint d\lambda\frac{W'_{\mbox{\scriptsize tree}}(\lambda)}
{(z-\lambda)(\lambda-b_-)^{\frac12}(\lambda-a_-)^{\frac12}
(\lambda-a_+)^{\frac12}(\lambda-b_+)^{\frac12}}.
\eeqa

The locations of the end points are determined so that $\omega(z)$ 
behaves like $\sim z^{-1}$ as $|z|\rightarrow\infty$. Defining 
\beqa
Q&=&\frac{a_+ +b_+ +a_- +b_-}4,
\eeqa
we compute the deviations of the end-point locations from 
 $a_\pm=\pm a$, $b_\pm = \pm b$ to first 
order in $Q$. The result is 
\beqa
a_\pm&=&\pm a +\left(1-\frac m{2\sqrt\mu}\right)Q ,\nonumber\\
b_\pm&=&\pm b +\left(1+\frac m{2\sqrt\mu}\right)Q,
\eeqa
where we have omitted the terms of $O(Q^2)$ (and will also do 
in the equations below).
$\omega(z)$ is modified from (\ref{omega(2-cut)}) to
\beqa
\omega(z)=\frac1{2\mu}\left(
mz + z^3 -(z+2Q)(z-b_-)^{\frac12}(z-a_-)^{\frac12}
(z-a_+)^{\frac12}(a-b_+)^{\frac12}
\right),
\eeqa
and hence the matrix model curve 
\beqa
y(z)=-(z+2Q)(z-b_-)^{\frac12}(z-a_-)^{\frac12}
(z-a_+)^{\frac12}(a-b_+)^{\frac12}.
\label{2-cutcurve}
\eeqa
Thus we find the $Q$-dependence of the periods of 
slightly asymmetric solutions as 
\beqa
S_\pm&=&\frac\mu2
\mp\frac{a^2bQ}\pi K\left(
\sqrt{1-\frac{a^2}{b^2}}
\right),
\label{S(2-cut)}\\
\Pi_\pm&=&
-\frac1{2\pi}\left(
W_{\mbox{\scriptsize tree}}(\Lambda_0)
+\frac \mu2 \log\frac \mu{\Lambda_0^4}-\frac \mu2 +\frac{m^2}4 
\right)
\mp\frac{a^2bQ}\pi K\left(\rule{0mm}{8mm}
\frac ab
\right)
\label{Pi(2-cut)}
\eeqa
to $O(Q)$, where 
\beqa
K(k)&=&\int_0^1\frac{dt}{\sqrt{(1-t^2)(1-k^2t^2)}}
\eeqa
is the complete elliptic integral of the first kind.
Using (\ref{S(2-cut)}) and (\ref{Pi(2-cut)}), we may readily find
in the symmetric limit ($Q=0$)
\beqa
\frac{\partial\Pi_\pm}{\partial S_\pm}
&=&-\frac{\partial\Pi_\pm}{\partial S_\mp}
~=~-\frac1{4\pi}\left(
\ln\frac\mu{\Lambda_0^4} \mp
{\textstyle \frac{K\left(\rule{0mm}{4mm}
\frac ab
\right)}
{K\left(
\sqrt{1-\frac {a^2}{b^2}}
\right)}}
\right).
\eeqa
Thus (\ref{Weffdef}) computes the effective superpotential as 
\beqa
W_{\mbox{\scriptsize eff}}(\hat S, w_\perp^\alpha)
&=&
N\left(
-\frac {\hat S}2\log\frac {\hat S}{\Lambda^4}+\frac {\hat S}2 -\frac{m^2}4 
\right)
\nonumber\\&&+
\left.\frac14
\left(
\log\frac\mu{\Lambda^4}
+{\textstyle \frac
{K\left(\rule{0mm}{4mm}
\frac ab
\right)}
{K\left(
\sqrt{1-\frac {a^2}{b^2}}
\right)}}
\right)\right|_{\mu=\hat S}
w_{\perp\alpha} w_\perp^\alpha.
\label{Weffphys}
\eeqa
where the $U(1)$ gauge superfield $w_\perp^\alpha$ has been 
defined by 
\beqa
S_\pm={\hat S}_\pm-\frac 1N w_{\pm\alpha}w_\pm^\alpha,~~~
w_\pm^\alpha =\frac12(w_0^\alpha \pm w_\perp^\alpha)
\eeqa
and the bare coupling term of the overall $U(1)$ $w_0^\alpha$ 
has been discarded in (\ref{Weffphys}). Similarly to the one-cut case, 
we have also defined the physical scale $\Lambda$ by the same equation as 
(\ref{scales}).
Minimizing  the first term with respect to $\hat S$, 
we find that this occurs when $\hat S = \Lambda^4$. 
Plugging this into (\ref{Weffphys}), we finally obtain 
\beqa
W_{\mbox{\scriptsize low}}
&=&N\left(
\frac{\Lambda^4}2 
-\frac{m^2}4
\right)
+\frac14
{\textstyle \frac
{K\left(\rule{0mm}{4mm}
\frac ab
\right)}
{K\left(
\sqrt{1-\frac {a^2}{b^2}}
\right)}}
w_{\perp\alpha} w_\perp^\alpha
~~~\mbox{(two-cut)},
\label{Wlow(2-cut)}
\eeqa
where $a$ and $b$ are given by the equations (\ref{ab}) with $\mu$ 
replaced by $\Lambda^4$. (\ref{Wlow(2-cut)}) agrees with the on-shell 
analysis of \cite{Ferrari2,Shih} derived using the matrix model curves with 
the double-zero factors removed.  

\subsection{The behavior near the transition point} 

\begin{figure}[t!]
\begin{center}
\rotatebox{90}{\hspace{25mm}$W_{\mbox{\scriptsize low}}/Nm^2$}
\resizebox{!}{6cm}{
\includegraphics{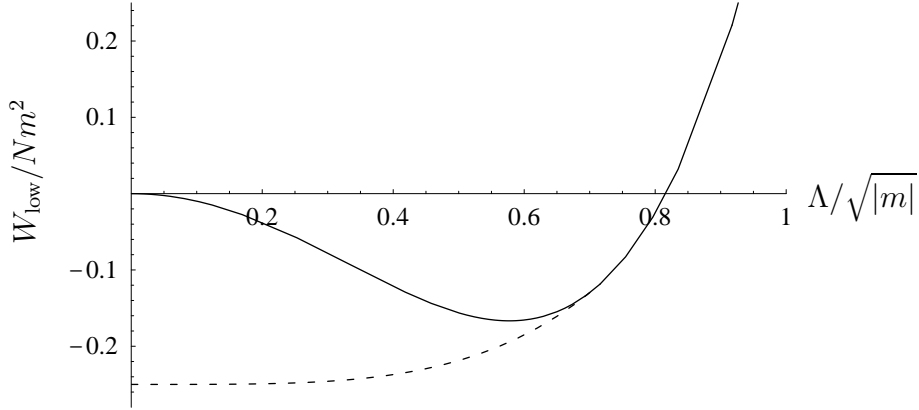}
}\raisebox{30mm}{$\Lambda/\sqrt{|m|}$}
\end{center}
\caption{Comparison of the minimum values of the effective 
superpotentials $W_{\mbox{\scriptsize low}}$. Below the critical scale 
$\Lambda=\sqrt{\frac{|m|}2}$, the broken phase (the two-cut solution, 
dashed line) is more favored than the unbroken phase 
(the one-cut solution, solid line). }
\label{W_low}
\end{figure}

Let us now compare (\ref{Wlow(1-cut)}) and (\ref{Wlow(2-cut)}) near 
the critical scale $\Lambda=\sqrt{\frac{|m|}2}$.
Figure \ref{W_low} shows comparison of the minimum values of the 
effective superpotentials $W_{\mbox{\scriptsize low}}$. 
No two-cut solution exists for  $\Lambda>\sqrt{\frac{|m|}2}$ and there are 
only one-cut solutions.
The values of the broken and the unbroken phases are smoothly connected 
at $\Lambda=\sqrt{\frac{|m|}2}$. 
Below the critical scale 
$\Lambda=\sqrt{\frac{|m|}2}$, the broken phase (two-cut solution, 
dashed line) is more favored than the unbroken phase (one-cut solution, 
solid line). This is what one would naively expect from the renormalization 
group argument: At high energies the $\mbox{Tr}\Phi^4$ term is more 
relevant than the $\mbox{Tr}\Phi^2$ term, enforcing the gauge group to be 
unbroken. At lower energies, the effect of the indentation of the potential 
becomes more relevant, and the gauge group is broken due to the Higgs 
mechanism. 
Since the effective superpotential in our model is proportional to a 
derivative of the free energy, the smoothness agrees with the fact that 
the transition is third order in the matrix model.

\begin{figure}[h!]
\begin{center}
\rotatebox{90}{\hspace{3cm}$i\tau_\perp$}
\resizebox{!}{6cm}{
\includegraphics{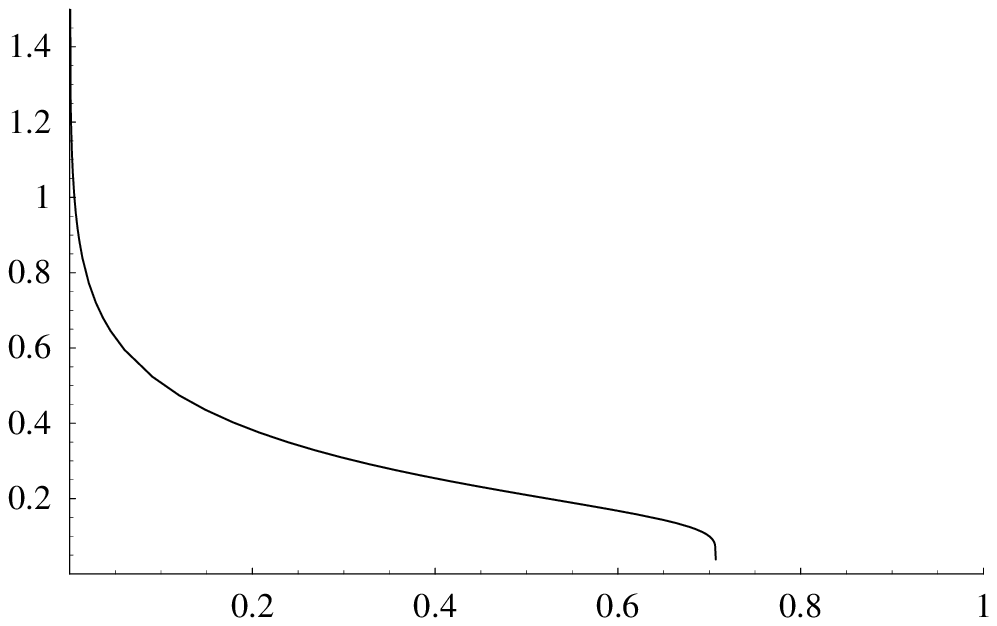}
}\\
\makebox{$\Lambda/\sqrt{|m|}$}
\end{center}
\caption{Scale dependence of the inverse square of the $U(1)$ coupling 
constant $i\tau_\perp\equiv\frac14
\frac
{K\left(\frac ab
\right)}
{K\left(
\sqrt{1-\frac {a^2}{b^2}}
\right)}
$.}
\label{Coeff}
\end{figure}

Figure \ref{Coeff} shows scale dependence of the inverse square of the 
$U(1)$ coupling constant 
$i\tau_\perp\equiv\frac14
\frac
{K\left(\frac ab
\right)}
{K\left(
\sqrt{1-\frac {a^2}{b^2}}
\right)}
$
(the coefficient of $w_{\perp\alpha}w_\perp^\alpha$) 
in the broken phase. It rapidly goes to zero like 
$\left(\log(\sqrt{\frac{|m|}2}-\Lambda)\right)^{-1}$
near $\Lambda=\sqrt{\frac{|m|}2}$, and 
the  kinetic term vanishes at the transition point; this is consistent with 
the fact that above $\Lambda=\sqrt{\frac{|m|}2}$ is the unbroken phase 
and no $U(1)$ gauge field is there.  Although the coupling constant grows 
very large near the transition point, it does not diverge until the parameters 
reaches the values where the one-cut solution starts. Therefore, one may say 
that the matrix model approach is still valid near the phase transition point, 
except right at the singularity. Note that the U(1) kinetic term does not 
vanish {\it off shell} at the transition point. These observations 
qualitatively agree with \cite{Shih} for the cubic-potential case.
At very low energies, $\tau_\perp$ is 
logarithmically divergent, reproducing the one-loop running as expected. 

\section{Painlev\'e II and Gravitational Corrections}
As we mentioned in the introduction, the singular behavior of the all-genus 
free energy, near the critical point we have considered in this paper, 
is known to be governed by the Painlev\'e II equation \cite{DSS} :
\beqa
\frac{d^2 O}{dx^2}+\frac12 xO-O^3=0,
\label{PII}
\eeqa
where $x$ is a variable related to the degree of polynomials
in the orthogonal polynomial method,
and $O(x)$ is related to the difference of the smooth limit of the even and 
odd recursion coefficients (See \cite{DSS} for more detail.).

In general, the orthogonal polynomial method 
extracts the leading critical singularity from the free energy, giving its 
expansion in terms of $\hat{N}^2(g-g_c)^{2-\gamma}$. 
The singular behavior of the free energy is controlled by $O(x)$ with 
large $x$.   
The solution of (\ref{PII}) is expanded around the infinity as
\beqa
O(x)&=&\sqrt{\frac x2}\left(
1-\frac1{4x^3}-\frac{73}{32 x^6}
-\frac{10657}{128 x^9}
%-\frac{13912277}{2048 x^{12}}
-\cdots
\right).
\eeqa
The leading singular behavior of the free energy is given by 
$\frac{d^2 F}{dz^2}\sim-(O(z))^2$ \cite{DSS},  
where $z^3=\hat{N}^{2}(1-\frac{4\hat S}{m^2})^3$.
Thus we find 
\beqa
F&\sim&\hat{N}^2 F_0 + F_1 + \left(
\frac{3\hat{N}^{-2}}{16 (1-\frac{4\hat S}{m^2})^3} 
+ \frac{63\hat{N}^{-4}}{32(1-\frac{4\hat S}{m^2})^6 } +\cdots
\right).
\eeqa
Since the string susceptibility $\gamma$ is  $-1$,  
the genus-0 and -1 terms are non-singular in the expansion and hence not 
reliable in this analysis, while the terms higher than $z^{-3}$ indicate large 
gravitational corrections \cite{DV3,OV1,OV2} near the transition point, 
Note that, although the matrix model size $\hat N$ is sent to infinity, 
the rank of the gauge group $N$ is {\em not} necessarily large in this 
double scaling limit.

\section{Conclusions}

We have used the classic solutions of a hermitian one-matrix model 
with an even quartic potential to compute low-energy effective superpotentials 
for ${\cal N}=1$, $SU(N)$ supersymmetric gauge theories. Since the solutions
are all $Z_2$ symmetric, the matrix models automatically satisfy the quantum 
tracelessness condition and describe a phase with the gauge group 
$SU(N/2)\times SU(N/2) \times U(1)$ (for an even $N$).

We have shown that the values of  the effective superpotentials are smoothly 
connected at the transition point, 
and the two-cut value of the superpotential is 
lower than that of the one-cut case below the critical scale.
The latter indicates that the broken phase is more favored at low energies, 
as naively expected.  At the transition point, the $U(1)$ coupling 
constant diverges, signaling the effect the light monopole, and  the $U(1)$ 
kinetic term consistently disappears there from the effective action, 
thereby confirming Ferrari's general formula for the critical 
behavior of the U(1) coupling constant.  We have also discussed that 
the Painlev\'e II equation of the double-scaled matrix model indicates 
large gravitational corrections near the transition point for the broken side,
if the gauge theory is coupled to a gravitational background.

In Ref.\cite{ES}, some evidence for a structure of the $N$-reduced KP 
hierarchy has been found in some ${\cal N}=1$ analogue of the 
Argyres-Douglas singularities. Our singular curve, on 
the other hand, does not belong to this class; nevertheless 
the all-genus free energy is described by the Painlev\'e II equation, 
which can be obtained as a similarity reduction of the modified KdV 
equation. It would be interesting to investigate if a structure of the
2-reduced KP (= KdV) hierarchy underlies our system in the sense 
of \cite{ES}. 

\section*{Acknowledgments}
We thank Yasuhiko Yamada for useful discussions.
The work of S.M. is supported in part by Grant-in-Aid
for Scientific Research (C)(2) \#14540286 from
The Ministry of Education, Culture, Sports, Science
and Technology.
%
%%%%%%%%%%%%%%%%%%
%  References
%%%%%%%%%%%%%%%%%%
%

\end{document}